\documentclass[12pt]{article}
\usepackage{amsmath}
\usepackage{graphicx}
\begin{document}
\baselineskip=18 pt
\begin{center}
{\large{\bf  Axially symmetric, asymptotically flat vacuum metric with a naked singularity and closed timelike curves}}
\end{center}

\vspace{.5cm}

\begin{center}
{\bf Debojit Sarma}\footnote{sarma.debojit@gmail.com},
{\bf Faizuddin Ahmed}\footnote{faizuddinahmed15@gmail.com}
and
{\bf Mahadev Patgiri}\footnote{mahadevpatgiri@mail.com}\\
 
{\it Department of Physics, Cotton College,}\\
{\it Guwahati-781001, India}
\end{center}

\vspace{.5cm}

\begin{abstract}

We present an axially symmetric, asymptotically flat empty space solution of the Einstein field equations containing a naked singularity.  The spacetime is regular everywhere except on the symmetry axis where it possess a true curvature singularity. The spacetime is of type D in the Petrov classification scheme and is locally isometric to the metrics of case IV in the  Kinnersley classification of type D vacuum metrics. Additionally, the spacetime also shows the evolution of closed timelike curves (CTCs) from an initial hypersurface free from CTCs.   

\end{abstract}



\vspace{.5cm}

\section{Introduction}

The weak cosmic censorship hypothesis (WCCH) of Penrose forbids naked singularities to be the endstate of gravitational collapse \cite{Pen1,Wald}. This statement, although accepted by many, has neither been proved nor disproved and, as a matter of fact, recent studies have shown that naked singularities can be formed in a continual gravitational collapse of a dust cloud beginning from a regular initial data \cite{Josh} indicating therefore the possibility of naked singularities occuring in nature. Since proving (or disproving) the WCCH has proved to be a difficult task, attempts have been made to provide the theoretical framework to devise a technique to distinguish between blackholes and naked singularities from astrophysical data mainly through gravitational lensing. Some significant work in this direction are the study of strong gravitational lensing in the Janis-Newman-Winicour spacetime \cite{Vir1,Vir2} and its rotating generalization \cite{Gyul} and notably (although not an exhaustive list) the work in \cite{Wern,Bam1,Bam2,Hio}. Other workers have shown that naked singularities and black holes may be differentiated by the properties of the accretion disks that accumulate around them (\cite{Chow} and references therein). Consequently, the study of naked singularities and spacetimes with such objects is of considerable current interest.

In \cite{Chow}, the authors have enumerated three possible end states of gravitational collapse. Of these, the last one, namely, that matter falls into the spacetime singularity preserving its symmetry structure and leading to a final configuration that is a vacuum spacetime with a singularity that is either naked, or covered, is of special interest to us.  In this paper, we attempt to construct such a spacetime -- a vacuum solution of Einstein's equations with a naked singularity. In what follows, we construct a Ricci flat metric with a singularity at $r=0$, $r$ being the radial coordinate and show that the spacetime is axially symmetric and asymptotically flat. In addition, we find that the metric admits closed timelike curves which appear at a certain moment implying that it also represents a time-machine spacetime.   

Closed timelike curves (CTCs), closed timelike geodesics (CTGs) and closed null geodesics (CNGs), which all imply violation of causality, are an intriguing aspect of the general theory of relativity and are present in certain solutions of Einstein's field equations. Beginning with G\"odel's universe \cite{Gode}, a considerable number of such causality violating solutions have been constructed. Van Stockum's solution \cite{Sto}, found years before G\"odel's, was later shown to possess CTCs \cite{Tip}. A few other examples of CTC spacetimes are Bonnor's rotating dust spacetime which has been analysed in \cite{Coll} and found to have CTCs, Gott's \cite{Gott} spacetime with two moving cosmic strings and wormhole spacetimes \cite{Morr}, a pure radiation metric with CTCs \cite{Sar1}. In fact, over the years, a number of solutions with CTCs have been obtained. CTCs have also been studied in the context of Brans-Dicke gravity \cite{Luis}, where rigid rotation of matter has been shown to generate these closed curves. Ori and co-workers \cite{Ori1,Ori2,Ori3,Ori4} have obtained solutions with CTCs which appear after a certain instant thereby raising the possibility of constructing a workable time machine at least in theory. We place spacetimes with CTCs (or CTGs or CNGs) into two categories. In the first, are the eternal time-machine spacetimes, such as God\"el's universe where CTCs pre-exist, while in the second, called true time-machines, CTCs appear at a certain moment.

Solutions with CTGs and CNGs have also been discussed in the literature. Soares found a class of cosmological solutions with rotating dust and electromagnetic fields admitting \cite{Soar}. Bonnor and Steadman \cite{Bon} have shown that a spacetime 
with two spinning particles, under special circumstances, allow for the existence of CTGs. 
A class of solutions of Einstein's field equations with CTGs representing spacetime outside a spinning cosmic string surrounded by a region of finite radial extension with vacuum energy  and a gas of strings \cite{Gro} have been recently found. Other examples are a cylindrically symmetric spacetime \cite{Gro1} which admit CTGs everywhere outside an axis and a vacuum spacetime with CNGs found quite recently \cite{Sar2}.

\section{Spacetime with a singularity} 
  
Consider the following line element, an empty spacetime, given by
\begin{eqnarray}
\nonumber
ds^2&=&-\sinh^2 r\cosh t\,\coth t\,dt^2+\cosh^2 r\,\sinh r\,dr^2+\mbox{csch}\; r\,dz^2 \nonumber\\
&+&\sinh^2 r\left(2\,\sqrt{2}\,\cosh t\,dt\,d\phi-\sinh t\,d\phi^2 \right )\quad
\label{4}
\end{eqnarray}
The coordinates are labelled $x^1=r$, $x^2=\phi$, $x^3=z$ and $x^4=t$ and their ranges are $r>0$, 
$-\infty < z < \infty$, $-\infty < t < \infty$ and $\phi$ is a periodic coordinate $\phi\sim\phi+\phi_0$, with $\phi_0>0$. The metric is Lorentzian with signature $(+,+,+,-)$ and the determinant of the corresponding metric tensor $g_{\mu\nu}$, $det\;g=-\cosh^2 r\,\sinh^4 r\,\cosh^2 t$. The metric (\ref{4}) is a solution of Einstein's equations in vacuum satisfying the equation 
$R_{\mu\,\nu}=0$. 

That the spacetime represented by (\ref{4}) is axisymmetric is clear from following. Consider the Killing vector ${\bf{\eta}}=\partial_{\phi}$ having the normal form
\begin{equation}
\eta^{\mu}=\left(0,1,0,0 \right )
\label{6}
\end{equation}
The corresponding co-vector is 
\begin{equation}
\eta_{\mu}=\sinh^2 r\left(0,-\sinh t,0,\sqrt{2}\,\cosh t\right )\quad
\label{7} 
\end{equation}
The vector (\ref{6}) satifies the Killing equation $\eta_{\mu\,;\,\nu}+\eta_{\nu\,;\,\mu}=0$. A spacetime is cyclically symmetric if it admits a Killing vector with spacelike, closed orbits. Axial symmetry means that the spacetime contains a non-empty axis of symmetry. The presence of this axis of symmetry is ensured if the norm of $\eta^{\mu}$ vanishes on the axis {\it i.e.} at $r=0$ (see \cite{Mars1,Mars2} and references therein). In our case we find that the norm is 
\begin{equation}
\eta_{\mu}\,\eta^{\mu}=-\,\sinh t\,\sinh^2 r\quad,
\label{8}
\end{equation}
which represents spacelike, closed orbits for $t<0$ and vanishes on the symmetry axis at 
$r\rightarrow 0$. The norm of $\eta_{\mu}$ changes sign for $t>0$ implying the formation of closed timelike curves which are the subject of the next section. 

The metric has a curvature singularity at $r=0$. We find that the Kretschmann scalar  (\ref{4}) 
\begin{equation}
K=R^{\mu\nu\rho\sigma}\,R_{\mu\nu\rho\sigma}=12\;{\mbox{csch}}^6 r
\label{9}
\end{equation}
and its differential invariant
\begin{equation}
R_{\mu\nu\rho\sigma\,;\,\lambda}\,R^{\mu\nu\rho\sigma\,;\,\lambda}=180\;{\mbox{csch}}^9 r
\label{inv}
\end{equation}
blow up at $r=0$ indicating a curvature singularity. We also note that both these scalar invariants approach zero rapidly as $r$ increases. 

The metric (\ref{4}) belongs to type D in the Petrov classification scheme. To show this, we construct the following set of null tetrads ${k,l,m,\bar{m}}$ for the metric (\ref{4}). Explicitly, these vectors have the form
\begin{equation}
k^{\mu}=\mbox{csch}\,r\left (0,\frac{1}{\sqrt{2}}\,\mbox{csch}\,t,0,\frac{1}{2}(2+\sqrt{2})\mbox{sech}\,t\right ) \quad,
\label{10}
\end{equation}
\begin{equation}
l^{\mu}=\mbox{csch}\,r\left (0,-\frac{1}{\sqrt{2}},0,\frac{1}{2}(-2+\sqrt{2})\mbox{tanh}\, t\right ) \quad, 
\label{11}
\end{equation}
\begin{equation}
m^{\mu}=\frac{1}{\sqrt{2\,\sinh r}}\,\left(\mbox{sech}\,r,0,i\,\sinh r,0\right )\quad,
\label{12}
\end{equation}
\begin{equation}
\bar{m}^{\mu}=\frac{1}{\sqrt{2\,\sinh r}}\,\left(\mbox{sech}\,r,0,-i\,\sinh r,0\right )\quad,
\label{13}
\end{equation}
where $i=\sqrt{-1}$. The set of null tetrads above is such that the metric tensor for the line element (\ref{4}) can be expressed as
\begin{equation}
g_{\mu \nu}=-k_{\mu}\,l_{\nu}-l_{\mu}\,k_{\nu}+m_{\mu}\,\bar{m}_{\nu}+\bar{m}_{\mu}\,m_{\nu} \quad .
\label{14}
\end{equation}
The vectors (\ref{10})---(\ref{13}) are null vectors and are orthogonal, except for $k_{\mu}l^{\mu}=-1$ and $m_{\mu}{\bar m}^{\mu}=1$. 

Using the set of null tetrads above, we find that, of the five Weyl scalars, only
\begin{equation}
\Psi_2=C_{\mu\nu\rho\sigma}\,k^{\mu}\,m^{\nu}\,{\bar m}^{\rho}\,l^{\sigma}=\frac{1}{2}{\mbox{csch}}^3 r
\label{16}
\end{equation}
is nonzero. The metric is clearly of type D in the Petrov classification scheme. It clear from (\ref{9}), (\ref{inv}) and (\ref{16}) that these scalars rapidly vanish as $r\rightarrow \infty$ leading to the conclusion that the metric is asymptotically flat. In fact, all nonzero curvature invariants for the spacetime have the form $A\; \mbox{csch}^n r$, $A$ being a real constant and $n$ a positive integer which means that their asymptotic behaviour is the same as that of the scalars in the equations (\ref{9}), (\ref{inv}) and (\ref{16}).

The vector $k^\mu$ is not only null but also geodesic, satisfying $k_{\mu;\nu}k^\nu=0$. We have calculated the kinematical properties of the metric such as expansion, shear and twist and they are given by
\begin{eqnarray}
\boldsymbol{\Theta}&=&\frac{1}{2}k^{\mu}_{\;\; ;\mu}=0 \\
\boldsymbol{\omega}^{2}&=&\frac{1}{2} (k_{\mu ;\nu} - k_{\nu ;\mu})k^{\mu ;\nu}=0\\
\boldsymbol{\sigma}\boldsymbol{{\bar\sigma}}&=&\frac{1}{2} (k_{\mu ;\nu} + k_{\nu ;\mu})k^{\mu ;\nu}-\boldsymbol{\Theta}^{2}=0
\end{eqnarray}
Hence the spacetime admits a expansion-free, twist-free and shear-free null geodesic congruence. 

All type D vacuum spacetimes are known and have been classified by Kinnersley \cite{Kinn}. A calculation of the Newman-Penrose spin co-efficients \cite{Steph} shows that for our case, the spin co-efficient $\rho=0$. Accordingly, the spacetime will be placed in case IV of the classification scheme in \cite{Kinn}. 

It is interesting property of the metric (\ref{4}) that it reduces to the Misner space \cite{Mis}
in two spacetime dimensions. To demonstrate this, we first apply the transformations $t\rightarrow \sinh ^{-1} t$\quad followed by\quad $ \phi\rightarrow \phi+\sqrt{2}\,\ln {t}$, to get 
\begin{equation}
ds^2=\cosh^2 r\,\sinh r\,dr^2+{\mbox{csch}}\,r\,dz^2+\sinh^2 r\,\left(-\,t\,d\phi^2+\frac{1}{t}\,dt^2 \right )\quad.
\label{17}
\end{equation}
A further transformation into (\ref{17}) by
\begin{equation}
\phi\rightarrow \phi+\ln {t}
\label{18}
\end{equation}
yields
\begin{equation}
ds^2=\cosh^2 r\,\sinh r\,dr^2+{\mbox{csch}}\,r\,dz^2-\sinh^2 r\,\left(2\,dt\,d\phi+t\,d\phi^2 \right )\quad
\label{20}
\end{equation}
The above reduces to the Misner space for constant $r$ and $z$ indicating that the metric discussed here is 4D extension of the Misner space.

\section{CTCs and Cosmic Time Machines}
 
A peculiar property of the spacetime (\ref{4}) is that it generates CTCs at a certain moment exhibiting time-machine like behaviour.  
Consider closed orbits of constant $r= r_0>0$, $z=z_0$ and $t=t_0$ where $r_0$, $z_0$ and $t_0$ are constants given by 1D line element
\begin{equation}
ds^2=g_{\phi\phi}\,d\phi^2=-\sinh t\,\sinh^2 r\,d\phi^2\quad
\label{1d}
\end{equation}
These orbits are null curves at $t=t_0=0$, spacelike throughout $t=t_0<0$, but become timelike {\it i.e.} $g_{\phi\phi}<0$ for $t=t_0>0$, which indicates the presence of CTCs. Hence CTCs form at an instant of time satisfying $t=t_0>0$. 

That these CTCs evolve from an initially spacelike $t=constant$ hypersurface can be determined by calculating the norm of the vector $\nabla_\mu t$ (or alternately from the value of $g^{tt}$ in the inverse metric tensor $g^{\mu\nu}$). A hypersurface $t=constant$ is spacelike when $g^{tt}<0$, timelike when $g^{tt}>0$ and null when $g^{tt}=0$. From the metric given in (\ref{4}),
\begin{equation}
g^{tt}=\sinh t\;\,{\mbox{sech}}^2\, t\;{\mbox{csch}}^2\, r
\label{5}
\end{equation}
Thus a hypersurface $t=constant$ is spacelike for $t<0$, timelike for $t>0$ and null at $t=0$. Thus the spacelike $t=constant<$0 hypersurface can be choosen as initial hypersurface over which initial data may be specified. There is a Cauchy horizon at $t=0$, called Chronology horizon, which separates the causal past and future in a past directed and a future directed manner. Hence the spacetime evolves from a partial Cauchy surface ({\it i.e.} Cauchy spacelike hypersurface) into a null hypersurface that is causally well-behaved up to a moment, {\it i.e.,} a null hypersurface $t=0$ and the formation of CTCs takes place from causally well-behaved initial conditions. This type of CTCs is different from those in the G\"odel universe \cite{Go}, where the CTCs pre-exist, and similar to those in \cite{Ori2}, where they form at some moment. The CTCs of the spacetime discussed in this paper is analogous to those formed in the Misner space \cite{Mis}. The metric for the Misner space in 2D 
\begin{equation}
ds_{Mis}^{2}=-\,2\,dT\,dX-T\,dX^2
\label{2}
\end{equation}
where $-\infty<T<\infty$ but the co-ordinate $X$ periodic. The curves $T=T_0$, where $T_0$ is a constant, are closed since $X$ is periodic. The curves $T<0$ are spacelike, $T>0$ are timelike, while the null curves $T=0$ form the chronology horizon. The second type of curves, namely, $T>0$ are closed timelike curves (CTCs).

An interesting possible interpretation of the metric (\ref{4}) is in terms of a cosmic time machine discussed by De Felice and co-workers. They have shown \cite{fel1} that if continued gravitational collapse leads to a strong curvature singularity, then strong causality is violated. This would mean that the entire spacetime becomes causally ill-behaved and a cosmic time machine is the result. They define a cosmic time machine as a spacetime which is asymptotically flat and admits closed non-spacelike curves \cite{fel2}. Our spacetime may represent such a cosmic time machine.

\vspace{1cm}
The authors declare that there is no conflict of interest regarding publication of this paper.

\end{document}